\documentclass[12pt]{article}
\usepackage{eurosym}
\usepackage{amsfonts}
\usepackage{amssymb}
\usepackage{amsmath}
\usepackage{graphicx}
\usepackage{color}
\usepackage{hyperref}

\numberwithin{equation}{section}

\begin{document}

\title{\textbf{Magnetic Mass in 4D AdS Gravity}}
\author{Ren\'{e} Araneda$^{1}$\thanks{%
renearanedag@hotmail.com}\thinspace , Rodrigo Aros$^{1}$\thanks{%
raros@unab.cl}\thinspace , Olivera Miskovic$^{2}$\thanks{%
olivera.miskovic@pucv.cl} \  \and and Rodrigo Olea$^{1}$\thanks{%
rodrigo.olea@unab.cl} \bigskip \\
{\small {$^{1}$}Departamento de Ciencias F\'{\i}sicas, Universidad Andres
Bello, Rep\'{u}blica 220, Santiago, Chile}\\
{\small {\ $^{2}$Instituto de F\'{\i}sica, Pontificia Universidad Cat\'{o}%
lica de Valpara\'{\i}so,}}\\
{\small {Casilla 4059, Valpara\'{\i}so, Chile}}}
\maketitle

\begin{abstract}
We provide a fully-covariant expression for the diffeomorphic charge in 4D
anti-de Sitter gravity, when the Gauss-Bonnet and Pontryagin terms are added
to the action. The couplings of these topological invariants are such that
the Weyl tensor and its dual appear in the on-shell variation of the action,
and such that the action is stationary for asymptotic (anti) self-dual
solutions in the Weyl tensor. In analogy with Euclidean electromagnetism,
whenever the self-duality condition is global, both the action and the total
charge are identically vanishing. Therefore, for such configurations the $%
\emph{magnetic}$ mass equals the Ashtekhar-Magnon-Das definition.
\end{abstract}

\section{Introduction}

Maxwell Lagrangian for electromagnetism is the simplest gauge-invariant
scalar that leads to second-order field equations. As it is well known,
gauge invariance is a consequence of using the Faraday tensor $F_{\mu \nu
}=\partial _{\mu }A_{\nu }-\partial _{\nu }A_{\mu }\,$, and not explicitly
the gauge connection $A_{\mu }$.

However, in four dimensions, Maxwell term is not the only Lagrangian
quadratic in $F$ that can be considered in an electromagnetism action. We
can always look at the physical implications which come from taking an
action of the form
\begin{equation}
I=-\frac{1}{4}\int\limits_{M}(F^{\mu \nu }F_{\mu \nu }+\gamma \,^{\ast
}F^{\mu \nu }F_{\mu \nu })\,dt\,d^{3}x\,,  \label{IEM}
\end{equation}%
where the second contribution is given in terms of the field strength and
its dual$~^{\ast }F^{\mu \nu }=\frac{1}{2}\epsilon ^{\mu \nu \alpha \beta
}F_{\alpha \beta }\,$, and it is called Pontryagin density
\begin{equation}
\mathcal{P}_{4}=\frac{1}{4}\,^{\ast }F^{\mu \nu }F_{\mu \nu }\,.
\label{P4covariant}
\end{equation}%
For a given real coupling constant $\gamma $, the second part of the action (%
\ref{IEM}) contributes just with a surface term, such that it does not alter
the bulk dynamics. Nevertheless, it may still modify the boundary conditions
in the variational problem and, eventually, the Noether current of the
theory.

In non-Abelian theories, Pontryagin is a topological term, which is added on
top of Yang-Mills Lagrangian with a pseudo-scalar coupling $\theta (x)$
(axion field) \cite{Wilczek}. This $\theta $-term is responsible for
violation of $CP$-symmetry in Quantum Chromodynamics.

In a more recent context, $\mathcal{P}_{4}$ has been considered to account
for properties of a new topological state in condensed matter physics known
as Topological Insulators \cite{TIs}.

In the Euclidean sector of the theory (\ref{IEM}), the electric field is
defined as $E_{i}=F_{0i}$, in terms of derivatives with respect to the Euclidean
time $x^{0}=it$ and the spatial coordinates $\{x^{i}\}$. In turn, the
magnetic field is the same as in the case of Lorentzian signature, that is, $%
B_{i}=\frac{1}{2}\epsilon ^{0ijk}F_{jk}$. With this in mind, the Pontryagin
invariant adopts the form%
\begin{equation}
\mathcal{P}_{4}=\mathbf{E\cdot B}\text{\thinspace },  \label{P4EB}
\end{equation}%
such that the Euclidean action $I^{E}=-iI$ reads%
\begin{equation}
I^{E}=\frac{1}{2}\int\limits_{M}\left( \mathbf{E}^{2}+\mathbf{B}^{2}+2\gamma
\,\mathbf{E\cdot B}\right) \,d^{4}x\,.  \label{IEEB}
\end{equation}

An arbitrary variation of this action produces
\begin{equation}
\delta I^{E}=\int\limits_{M}\left( \partial _{\mu }F^{\mu \nu }+\gamma
\,\partial _{\mu }\,^{\ast }F^{\mu \nu }\right) \delta A_{\nu
}\,d^{4}x-\int\limits_{\partial M}\left( F^{\mu \nu }+\gamma \,^{\ast
}F^{\mu \nu }\right) \delta A_{\nu }\,d\Sigma _{\mu }\,,  \label{varIE}
\end{equation}%
where the bulk integral yields Maxwell equation and second term which
vanishes due to Bianchi identity, $\partial _{\mu }\,^{\ast }F^{\mu \nu }=0$.

In order to have a well-defined action principle ($\delta I^{E}=0$), it is
necessary that field equations hold and that the surface term vanishes for a
given boundary condition. Usually one fixes the vector potential on the
boundary, i.e., $\delta A_{\mu }=0$. In particular, when the boundary is a
surface separating two regions of the space, this Dirichlet condition
defines the junction conditions for the electric and magnetic fields across
the surface in terms of sources present on the boundary, i.e., the surface
charge and current densities.

Another way to achieve a well-posed variational principle is to demand that
an asymptotic (anti) self-duality condition holds at the boundary, that is,
\begin{equation}
F^{\mu \nu }=\pm ^{\ast }F^{\mu \nu }\text{ \quad at }\partial M\,,
\end{equation}%
such that this argument fixes the Pontryagin coupling as $\gamma =\mp 1$.

Self-duality is a global symmetry of the sourceless Maxwell equation, where the
electric and magnetic degrees of freedom are interchanged. An extension to
electromagnetism with sources should necessarily include a \emph{magnetic }%
charge. In the Hamiltonian formulation of Maxwell theory, self-duality is an
off-shell symmetry, as shown by Deser and Teitelboim in Ref.\cite%
{Deser-Teitelboim76}.

Using the identity%
\begin{equation}
F_{\mu \nu }F^{\mu \nu }=\frac{1}{2}\,(F_{\mu \nu }F^{\mu \nu }+\,^{\ast
}F_{\mu \nu }\,^{\ast }F^{\mu \nu })\,,
\end{equation}%
the Euclidean action can be rewritten as%
\begin{equation}
I^{E}=\frac{1}{8}\int\limits_{M}(F^{\mu \nu }\mp \,^{\ast }F^{\mu \nu
})^{2}\,d^{4}x\,.
\end{equation}%
It is worth noticing that for a global (anti) self-duality condition, the
action is identically zero. The solutions in this case are known as
Euclidean Instantons. The condition $I^{E}=0$ defines a number of ground
states of the theory, where $F_{\mu \nu }=0$ is the simplest case of
globally self-dual solution.

Invariance under a $U(1)$ gauge transformation, where the gauge field
changes as $\delta _{\lambda }A_{\nu }=\partial _{\nu }\lambda $, leads to a
conservation law associated to this symmetry. Indeed, using the general
on-shell variation of the action (\ref{varIE}) in the Noether theorem (see
Appendix \ref{Noether}), a conserved charge can be constructed,
\begin{equation}
Q[\lambda ]=-\int\limits_{S^{2}}\left( F^{\mu \nu }\mp \,^{\ast }F^{\mu \nu
}\right) \lambda \,d\Sigma _{\mu \nu }\,,  \label{Qlambda}
\end{equation}%
where $d\Sigma _{\mu \nu }$ is the dual of the infinitesimal surface element
in $S^{2}$. Since the gauge parameter $\lambda $ is covariantly constant in
the asymptotic region, it can be normalized as $\lambda =1$. The first term
is the contribution due to the Noether current for Maxwell electromagnetism,
$J^{\mu }[\lambda ]\sim F^{\mu \nu }\partial _{\nu }\lambda \,$,
the conservation of which produces the electric charge. The second term is derived from a
topological current, $\tilde{J}^{\mu }[\lambda ]\sim \,^{\ast }F^{\mu \nu
}\partial _{\nu }\lambda \,$, and it corresponds to the magnetic flux across
the sphere $S^{2}$, i.e., magnetic charge \cite{Nakahara}. Simply put, Eq.(%
\ref{Qlambda}) identifies the Noether charge obtained from a topological
term with a topological charge derived from Bianchi identity.

It is evident from the formula (\ref{Qlambda}) that any globally (anti)
self-dual solution will have a vanishing charge. This argument reinforces
the idea that such a configuration can be regarded as a ground state of the
theory, and provides a firmer ground for the extension of self-duality
condition to AdS gravity discussed below.

\section{4D AdS Gravity and Pontryagin Invariant}

The addition of topological invariants, which modify the boundary dynamics
of AdS gravity, was considered more than fifteen years ago in Refs. \cite%
{Aros:1999id,Aros:1999kt}. Indeed, the regulation of the Noether current by
the addition of the Euler density provides a generic expression for the mass
and other charges for even-dimensional asymptotically AdS (AAdS) spaces. As
this procedure was performed in first-order formalism, its relation to other
approaches was not clear at that moment, even though the equivalence to
Hamiltonian charges was given in Ref. \cite{Aros:2001gz}. In particular its
relevance within the framework of anti-de Sitter/Conformal Field Theory
(AdS/CFT) correspondence \cite{Witten:1998qj} was certainly unknown.
However, this approach was later translated into metric formalism in Ref.%
\cite{Olea2n}, and understood as the addition of counterterms which depend
on the extrinsic curvature. It was then extended to odd dimensions \cite%
{Olea:2006vd}, giving rise to an alternative regularization scheme known as
\emph{Kounterterms}. Furthermore, the connection to Holographic
Renormalization \cite{Skenderis:2002wp} in the context of AdS/CFT
correspondence was shown in Refs.\cite{Miskovic:2009bm,MOT}, as the
asymptotic expansion of the extrinsic curvature reproduces the standard
counterterm series \cite{EJM,KLS}.

The simplest example of regularization using topological invariants is the
addition of Gauss-Bonnet term to four-dimensional AdS action studied in Ref.%
\cite{Aros:1999id},
\begin{equation}
I_{4}=\frac{1}{16\pi G}\int\limits_{M}d^{4}x\sqrt{g}\left[ R+\frac{6}{\ell
^{2}}+\frac{\ell ^{2}}{4}\,\left( R_{\mu \nu \alpha \beta }R^{\mu \nu \alpha
\beta }-4R_{\mu \nu }R^{\mu \nu }+R^{2}\right) \right] ,  \label{I4}
\end{equation}%
where $\ell $ is the AdS radius and $g=\left\vert \det (g_{\mu \nu
})\right\vert $. This is the same as the quadratic action given by MacDowell
and Mansouri in four dimensions in Ref.\cite{MacDowell-Mansouri} (see also
Ref. \cite{Garcia-Compean}), which was later extended to higher dimensions
by Vasiliev \cite{Vasiliev}.

The Gauss-Bonnet coupling is such that the action is stationary for
asymptotically locally AdS spaces, where the spacetime curvature tends to a
constant, i.e., $R_{\alpha \beta }^{\mu \nu }\rightarrow -\frac{1}{\ell ^{2}}%
\delta _{\lbrack \alpha \beta ]}^{[\mu \nu ]}$. This is evident from the
on-shell variation of $I_{4}$,
\begin{equation}
\delta I_{4}=\frac{\ell ^{2}}{64\pi G}\int\limits_{\partial M}d^{3}x\,\sqrt{h%
}\,n_{\mu _{1}}\delta _{\lbrack \nu _{1}\nu _{2}\nu _{3}\nu _{4}]}^{[\mu
_{1}\mu _{2}\mu _{3}\mu _{4}]}\,g^{\nu _{2}\gamma }\delta \Gamma _{\gamma
\mu _{2}}^{\nu _{1}}\left( R_{\mu _{3}\mu _{4}}^{\nu _{3}\nu _{4}}+\frac{1}{%
\ell ^{2}}\,\delta ^{\lbrack \nu _{3}\nu _{4}]}_{[\mu _{3}\mu _{4}]}\right)
\,,  \label{deltaI4}
\end{equation}%
where $n_{\mu _{1}}$ is an outward pointing unit normal to the boundary with
the induced metric $h_{ij}$, and $h=\left\vert \det (h_{ij})\right\vert $.
Also, $\delta _{\lbrack \nu _{1}\nu _{2}\nu _{3}\nu _{4}]}^{[\mu _{1}\mu
_{2}\mu _{3}\mu _{4}]}$ is the totally anti-symmetric Kronecker delta
defined as $\det \left[ \delta _{\nu _{1}}^{\mu _{1}}\cdots \delta _{\nu
_{4}}^{\mu _{4}}\right] $. The key argument that supports the finiteness of
the action principle is given by the fact that, for any solution of the
Einstein equation $R_{\mu \nu }=-\frac{3}{\ell ^{2}}g_{\mu \nu }$, the Weyl
tensor is%
\begin{equation}
W_{\mu \nu }^{\alpha \beta }=R_{\mu \nu }^{\alpha \beta }+\frac{1}{\ell ^{2}}%
\,\delta _{\lbrack \mu \nu ]}^{[\alpha \beta ]}\,,  \label{WRdelta}
\end{equation}
which is exactly the quantity that appears at the right hand side of Eq.(\ref%
{deltaI4}). The Weyl tensor is the only combination between the Riemann and
Ricci tensors that has a suitable asymptotic behavior. A formal proof of the
finiteness of the action, however, requires precise fall-off conditions in
the metric, valid for any AAdS spacetime \cite{Miskovic:2009bm}.

The appearance of the Weyl tensor in the surface term coming from the
variation of the total action (\ref{I4}) reflects the link to Conformal Mass
definition in AAdS gravity \cite{Ashtekar-Magnon-Das}. Indeed, upon suitable
expansion of the tensors involved, one can prove that the physical
information on the conformal boundary is encoded in the electric part of the
Weyl tensor \cite{Miskovic:2009bm,JKMO}.

Gauss-Bonnet is not the only possible topological invariant for the Lorentz
group one can construct in four dimensions. Indeed, Pontryagin density in
gravity \cite{Deser}, where the Riemann tensor plays the role of the field strength in Eq.(%
\ref{P4covariant}), is given by
\begin{equation}
\mathcal{P}_{4}=-\frac{1}{4}\,\epsilon ^{\mu \nu \alpha \beta }R_{\mu \nu
}^{\sigma \lambda }R_{\sigma \lambda \alpha \beta }\text{ }.
\end{equation}%
As the Pontryagin is a closed form, it can be written locally as the
divergence of a Chern-Simons density current%
\begin{equation}
\mathcal{P}_{4}=\partial _{\mu }\left[ \epsilon ^{\mu \nu \alpha \beta
}\left( \Gamma _{\nu \lambda }^{\sigma }\partial _{\alpha }\Gamma _{\beta
\sigma }^{\lambda }+\frac{2}{3}\Gamma _{\nu \lambda }^{\sigma }\Gamma
_{\alpha \epsilon }^{\lambda }\Gamma _{\beta \sigma }^{\epsilon }\right) %
\right] .
\end{equation}

We consider the addition of the Pontryagin density on top of a finite AdS
action given by Euclideanized version of Eq.(\ref{I4}), that is,
\begin{equation}
I=I_{4}+\frac{\ell ^{2}}{32\pi G}\,\gamma \int\limits_{M}d^{4}x\,\mathcal{P}%
_{4},  \label{I+P4}
\end{equation}%
where $\gamma $ is a coupling constant yet to be determined. We emphasize
the fact that, in this case, $\gamma $ is a given constant, not a function.
As a consequence, the action in Eq.(\ref{I+P4}) does not describe the
Chern-Simons modified gravity theory developed by R. Jackiw and S.-Y. Pi in
Ref.\cite{Jackiw:2003pm}, where by analogy to dynamic couplings of
electromagnetic Pontryagin, one is able to modify the gravitational field
equation in the bulk.

It is direct to check that the addition of the Pontryagin density does not
introduce divergences when evaluating AAdS solutions. Indeed, $\mathcal{P}%
_{4}$ is zero for AdS black holes and, at most, finite for gravitational
instantons, as it will be discussed below.

The addition of $\mathcal{P}_{4}$ produces a new surface term with respect to the
one in Eq.(\ref{deltaI4}), which is proportional to the dual of the Riemann
tensor, i.e.,%
\begin{equation}
\delta I=\frac{\ell ^{2}}{64\pi G}\int\limits_{\partial {M}}d^{3}x\,\sqrt{h}%
\,n_{\mu _{1}}\delta _{\lbrack \nu _{1}\nu _{2}\nu _{3}\nu _{4}]}^{[\mu
_{1}\mu _{2}\mu _{3}\mu _{4}]}\,g^{\nu _{2}\gamma }\,\delta \Gamma _{\gamma
\mu _{2}}^{\nu _{1}}\left( W_{\mu _{3}\mu _{4}}^{\nu _{3}\nu _{4}}-\frac{%
\gamma }{2\sqrt{g}}\,\epsilon ^{\nu _{3}\nu _{4}\alpha \beta }R_{\alpha
\beta \mu _{3}\mu _{4}}\right) .  \label{W+betadualR}
\end{equation}%
It is adequate to perform a shift in the curvature of the type $R_{\alpha
\beta \mu _{3}\mu _{4}}\rightarrow R_{\alpha \beta \mu _{3}\mu _{4}}+\frac{1%
}{\ell ^{2}}\left( g_{\alpha \mu _{3}}g_{\beta \mu _{4}}-g_{\beta \mu
_{3}}g_{\alpha \mu _{4}}\right) $, as the second term is identically zero
due to the symmetry in the indices. In doing so, the variation of the total
action can be rewritten as
\begin{equation}
\delta I=\frac{\ell ^{2}}{64\pi G}\int\limits_{\partial {M}}d^{3}x\,\sqrt{h}%
\,n_{\mu _{1}}\delta _{\lbrack \nu _{1}\nu _{2}\nu _{3}\nu _{4}]}^{[\mu
_{1}\mu _{2}\mu _{3}\mu _{4}]}\,g^{\nu _{2}\gamma }\delta \Gamma _{\gamma
\mu _{2}}^{\nu _{1}}\left( W_{\mu _{3}\mu _{4}}^{\nu _{3}\nu _{4}}-\gamma
\,^{\ast }W_{\mu _{3}\mu _{4}}^{\nu _{3}\nu _{4}}\right) \,,
\end{equation}%
in terms of the dual of the Weyl tensor
\begin{equation}
^{\ast }W_{\alpha \beta \mu \nu }=\frac{1}{2}\,\sqrt{g}\,\epsilon _{\alpha
\beta \sigma \lambda }W_{\mu \nu }^{\sigma \lambda }\,.
\end{equation}

By analogy to the EM case, one can determine $\gamma $ by demanding an
asymptotic (anti)-self duality condition on the Weyl tensor,
\begin{equation}
W_{\alpha \beta \mu \nu }=\pm \,^{\ast }W_{\alpha \beta \mu \nu }\,.
\label{selfdualWeyl}
\end{equation}%
The action is truly stationary if the field equations hold in the bulk and
the surface term vanishes at the boundary. Therefore, a well-defined action
principle for the boundary condition (\ref{selfdualWeyl}) implies that the
Pontryagin coupling is $\gamma =\pm 1$ \cite{Miskovic:2009bm}.

As the Weyl tensor carries information on the normalizable modes in AdS
gravity, the above condition implies a nontrivial relation between different
components of the Weyl tensor at a holographic order. Indeed, asymptotic
self-duality for the Weyl tensor, which appears naturally at the boundary
when one adds Gauss-Bonnet (parity-preserving) and Pontryagin
(parity-violating) topological invariants, seems to be the ultimate reason
behind holographic stress tensor/Cotton tensor duality, which arises when
dealing with AdS instantons \cite{deHaro}, hydrodynamic perturbations around
AdS$_{4}$ black holes \cite{Bakas} and electric/magnetic duality in
Riemann-Cartan-AdS gravity \cite{Mansi et al}.

Only for the particular value of the Pontryagin coupling discussed above,
the on-shell action adopts the compact form \cite{Miskovic:2009bm}
\begin{equation}
I=\frac{\ell ^{2}}{512\pi G}\int_{M}d^{4}x\,\sqrt{g}\,\delta _{\lbrack \nu
_{1}\nu _{2}\nu _{3}\nu _{4}]}^{[\mu _{1}\mu _{2}\mu _{3}\mu _{4}]}\left(
W_{\mu _{1}\mu _{2}}^{\nu _{1}\nu _{2}}\pm 
\,^*W^{\nu _{1}\nu_{2}}_{\mu _{1}\mu _{2}}\right) \left( W_{\mu _{3}\mu _{4}}^{\nu _{3}\nu
_{4}}\pm \,^{\ast }W^{\nu _{3}\nu _{4}}_{\mu _{3}\mu _{4}}\right) \,,
\end{equation}%
in terms of the Weyl tensor and its dual, where we have used the identities%
\begin{equation}
^{\ast }W_{\nu _{1}\nu _{2}}^{\mu _{1}\mu _{2}}=\frac{1}{4}\,\delta
_{\lbrack \nu _{1}\nu _{2}\nu _{3}\nu _{4}]}^{[\mu _{1}\mu _{2}\mu _{3}\mu
_{4}]}\,^{\ast }W^{\nu _{3}\nu _{4}}_{\mu _{3}\mu _{4}}\,,
\end{equation}%
and
\begin{equation}
\delta _{\lbrack \nu _{1}\nu _{2}\nu _{3}\nu _{4}]}^{[\mu _{1}\mu _{2}\mu
_{3}\mu _{4}]}\,^{\ast }W^{\nu _{1}\nu _{2}}_{\mu _{1}\mu _{2}}\,^{\ast
}W^{\nu _{3}\nu _{4}}_{\mu _{3}\mu _{4}}=\delta _{\lbrack \nu _{1}\nu
_{2}\nu _{3}\nu _{4}]}^{[\mu _{1}\mu _{2}\mu _{3}\mu _{4}]}\,W^{\nu _{1}\nu
_{2}}_{\mu _{1}\mu _{2}}\,W^{\nu _{3}\nu _{4}}_{\mu _{3}\mu _{4}}\,.
\end{equation}%
This action has been recently studied in the context of a search for a
pure-spin connection formulation for General Relativity \cite%
{Basile-Bekaert-Boulanger}.

In what follows, we compute the Noether charges for the gravity action $I$
using Wald's method \cite{IyerandWald,Iyer:1995kg}. This is the
fully-covariant version of the boundary derivation which associates the
addition of the Gauss-Bonnet term to the \emph{electric }part of the Weyl
tensor and the addition of Pontryagin to \emph{magnetic} part of the Weyl
tensor (see Appendix \ref{AAdS}).

\section{Covariant Noether charges and Topological Invariants}

Noether theorem provides a conserved current $J^{\mu }$ ($\partial _{\mu }(%
\sqrt{g}J^{\mu })=0)$, for a given symmetry of an action. Indeed, global
isometries in gravitational solutions imply the existence of a Noether
charge defined as
\begin{equation}
Q=\int\limits_{\partial M}d^{3}x\,\sqrt{h}\,n_{\mu }J^{\mu }.
\end{equation}%
When $J^{\mu }$ can be globally written as $J^{\mu }=\partial _{\nu }(\sqrt{h%
}Q^{\mu \nu })$ in $\partial M$, the Noether charge can be expressed as an
integral on the two-dimensional surface $\partial \Sigma $ with the metric $%
\sigma _{mn}$ and $\sigma =\left\vert \det (\sigma _{mn})\right\vert $,%
\begin{equation}
Q=\int\limits_{\partial \Sigma }\sqrt{\sigma }\,d^{2}x\,n_{\mu }u_{\nu
}\,Q^{\mu \nu },
\end{equation}%
where $u_{\nu }$ is a unit timelike vector, normal at every point to $\Sigma
$ (see Appendix \ref{Noether}).

For the case under study here, we follow Wald's procedure defined in Refs.
\cite{IyerandWald,Iyer:1995kg}, which allows us to construct the Noether
charges in an arbitrary gravity theory. We consider a Lagrangian density $L$%
, which depends on the metric, curvature and covariant derivatives of the
curvature,
\begin{eqnarray}
L &=&L(g_{\mu \nu },R_{\mu \nu \alpha \beta },\nabla _{\gamma _{1}}R_{\mu
\nu \alpha \beta },\cdots  \notag \\
&&\cdots ,\nabla _{(\gamma _{1}\cdots }\nabla _{\gamma _{m})}R_{\mu \nu
\alpha \beta },\psi ,\nabla _{\gamma _{1}}\psi ,\nabla _{(\gamma _{1}\cdots
\gamma _{l})}\psi )\,.
\end{eqnarray}%
One can also include matter fields, collectively denoted by $\psi $, and
derivatives of them.

For this general class of theories, the conserved current corresponding to a
set of Killing vectors $\{\xi ^{\mu }\}$, is given by the expression (see
Appendix \ref{Noether})%
\begin{equation}
\sqrt{g}J^{\mu }=\Theta ^{\mu }(\delta _{\xi }\Gamma )+\Theta ^{\mu }(\delta
_{\xi }g)+\xi ^{\mu }L,
\end{equation}%
assuming that the surface term $\Theta ^{\mu }$ can be split in a part that
contains variations of the Christoffel symbol and another that contains
variations of the metric tensor. Due to the fact that $\Theta ^{\mu }(\delta
_{\xi }g)$ is proportional to the Lie derivative of the metric, using the
Killing equation, this term can be set to zero. Then, the conserved current
adopts the form%
\begin{equation}
\sqrt{g}J^{\mu }=2E^{\mu \nu \alpha \beta }g_{\alpha _{\lambda }}\delta
_{\xi }\Gamma _{\nu \beta }^{\lambda }+\xi ^{\mu }L,  \label{JmuE}
\end{equation}%
where $E^{\mu \nu \alpha \beta }$ is the variation of $L$ with respect to
the Riemann tensor $R_{\mu \nu \alpha \beta }\,$. The diffeomorphism
transformation of the Christoffel symbol is given by%
\begin{eqnarray}
\delta _{\xi }\Gamma _{\nu \beta }^{\lambda } &=&-\frac{1}{2}\,g^{\lambda
\rho }\left( \nabla _{\beta }\pounds _{\xi }g_{\rho \nu }+\nabla _{\nu }%
\pounds _{\xi }g_{\rho \beta }-\nabla _{\rho }\pounds _{\xi }g_{\beta \nu
}\right)  \notag \\
&=&-\frac{1}{2}\,\left( \nabla _{\nu }\nabla _{\beta }\xi ^{\lambda }+\nabla
_{\beta }\nabla _{\nu }\xi ^{\lambda }\right) +\frac{1}{2}\,\left( R_{~\beta
\nu \sigma }^{\lambda }+R_{\ \nu \beta \sigma }^{\lambda }\right) \xi
^{\sigma }\,,  \label{LieChristoffel}
\end{eqnarray}%
which produces a current%
\begin{equation}
\sqrt{g}J^{\mu }=-E^{\mu \nu \alpha \beta }\left[ 2\nabla _{\nu }\nabla
_{\beta }\xi _{\alpha }-\left( R_{\alpha \beta \nu \sigma }+2R_{\alpha \nu
\beta \sigma }\right) \xi ^{\sigma }\right] +\xi ^{\mu }L\,,
\end{equation}%
where we have used the identity that involves the commutator of two
covariant derivatives,%
\begin{equation}
\lbrack \nabla _{\beta },\nabla _{\nu }]\xi _{\alpha }=R_{\beta \nu \alpha
\sigma }\xi ^{\sigma }\,.
\end{equation}%
A minor arrangement can be performed in the above expression for the
current, as the tensor $E^{\mu \nu \alpha \beta }$ inherits a given symmetry
in the indices which is derived from first Bianchi identity, that is,
\begin{equation}
0=R_{\alpha \beta \nu \sigma }+R_{\beta \nu \alpha \sigma }+R_{\nu \alpha
\beta \sigma }\,,
\end{equation}%
which implies
\begin{equation}
E^{\mu \nu \alpha \beta }\left( R_{\alpha \beta \nu \sigma }-2R_{\alpha \nu
\beta \sigma }\right) =0\,.
\end{equation}%
Finally, the formula for the Noether current in a generic gravity theory is
given by%
\begin{equation}
\sqrt{g}J^{\mu }=2E^{\mu \nu \alpha \beta }\left( \nabla _{\nu }\nabla
_{\alpha }\xi _{\beta }+R_{\alpha \beta \nu \sigma }\xi ^{\sigma }\right)
+\xi ^{\mu }L\,.
\end{equation}%
For the case under study, we can see that the Noether current associated to
the EH Lagrangian plus GB term in Eq.(\ref{I4}) yields
\begin{equation}
J_{4}^{\mu }[\xi ]=\frac{\ell ^{2}}{64\pi G}\,\delta _{\lbrack \alpha \beta
\gamma \delta ]}^{[\mu \nu \lambda \sigma ]}\,W_{\lambda \sigma }^{\gamma
\delta }\nabla _{\nu }\nabla ^{\alpha }\xi ^{\beta }
\end{equation}%
which, using the second Bianchi identity in the indices $\nu \lambda \sigma $%
, can be written down as a total derivative,%
\begin{equation}
J_{4}^{\mu }[\xi ]=\frac{\ell ^{2}}{64\pi G}\,\nabla _{\nu }\left( \delta
_{\lbrack \alpha \beta \gamma \delta ]}^{[\mu \nu \lambda \sigma
]}W_{\lambda \sigma }^{\gamma \delta }\nabla ^{\alpha }\xi ^{\beta }\right)
\,.
\end{equation}%
Here, we have used the field equations and permutational identities in order
to eliminate additional terms in the curvature, which are coming from the
Lie derivative acting on the Christoffel symbol (\ref{LieChristoffel}).
Integrated on $\partial \Sigma $\thinspace , the above expression produces
the charge
\begin{equation}
Q_{4}[\xi ]=\frac{\ell ^{2}}{64\pi G}\int\limits_{\partial \Sigma }d^{2}x\,%
\sqrt{\sigma }n_{\mu }u_{\nu }\,\delta _{\lbrack \alpha \beta \gamma \delta
]}^{[\mu \nu \lambda \sigma ]}\,\nabla ^{\alpha }\xi ^{\beta }W_{\lambda
\sigma }^{\gamma \delta }\,.  \label{n=7}
\end{equation}

Taking now the Lagrangian density corresponding to the Pontryagin term, we
have%
\begin{equation}
E_{\mathcal{P}_{4}}^{\mu \nu \alpha \beta }=\mp \frac{\ell ^{2}}{64\pi G}%
\,\epsilon ^{\mu \nu \lambda \sigma }R_{\lambda \sigma }^{\alpha \beta }\,,
\end{equation}%
expression which determines the current associated to this term as
\begin{equation}
J_{\mathcal{P}_{4}}^{\mu }=\mp \frac{\ell ^{2}}{64\pi G}\,\delta _{\lbrack
\alpha \beta \gamma \delta ]}^{[\mu \nu \lambda \sigma ]}\,\nabla _{\nu
}\left( \nabla ^{\alpha }\xi ^{\beta }\,^{\ast }W_{\lambda \sigma }^{\gamma
\delta }\right) .  \label{P4Current}
\end{equation}%
As a consequence, the total Noether charge computed for the AdS gravity
action with the addition of topological invariants is
\begin{equation}
Q[\xi ]=\frac{\ell ^{2}}{64\pi G}\int\limits_{\partial \Sigma }d^{2}x\,\sqrt{%
\sigma }\,n_{\mu }u_{\nu }\delta _{\lbrack \alpha \beta \gamma \delta
]}^{[\mu \nu \lambda \sigma ]}\,\nabla ^{\alpha }\xi ^{\beta }\left(
W_{\lambda \sigma }^{\gamma \delta }\mp \,^{\ast }W_{\lambda \sigma
}^{\gamma \delta }\right) \,.  \label{SelfDualNoetherCharge}
\end{equation}%
It is then that the analogy with self-dual electromagnetism becomes evident:
self-dual or anti self-dual solutions in AdS gravity have mass (and other
conserved quantities) identically zero. Such a configuration is
a vacuum state, which reaches a minimum of the Euclidean action.

\subsection{Taub-NUT/Bolt AdS solutions}

For static black hole and even Kerr-AdS solutions, the magnetic part of the
Weyl tensor is zero, such that there is no contribution to the current (\ref%
{P4Current}). Therefore, non-trivial examples to evaluate the above
expressions for the conserved quantities are Taub-NUT and Taub-Bolt AdS
solutions. These spaces are Euclidean gravitational solutions to the
Einstein equations characterized by a line element \cite%
{N.U.T.,PagePope,KramerStephani}
\begin{equation}
ds^{2}=f(r)\left( d\tau +2n\cos \theta \,d\phi \right) ^{2}+\frac{dr^{2}}{%
f(r)}+\left( r^{2}-n^{2}\right) \left( d\theta ^{2}+\sin ^{2}\theta \,d\phi
^{2}\right) \,,
\end{equation}%
where the function $f(r)$ is given by ($G=1$)
\begin{equation}
f(r)=\frac{r^{2}-2Mr+n^{2}-\frac{3}{\ell ^{2}}\left( n^{4}+2n^{2}r^{2}-\frac{%
r^{4}}{3}\right) }{r^{2}-n^{2}}\,.
\end{equation}%
Here, $n$ is a parameter, and $M$ is identified as the solution mass \cite%
{Aros:1999id,Hawking:1998ct}. The Taub-NUT-AdS solution is defined by the
condition $f(|n|)=0$, but one still has to eliminate the conical
singularities that appear at $r=|n|$. By imposing a regularity condition,
which is given by $f^{\prime }(|n|)=1/2n$, the electric mass takes the
particular value%
\begin{equation}
Q_{4}^{\mathrm{NUT}}[\partial _{\tau }]=M_{\mathrm{NUT}}=\pm n\left( 1-4\ell
^{-2}n^{2}\right) \,.
\end{equation}%
This value of $M$ is the exact point where the Weyl tensor becomes globally
(anti) self-dual \cite{Atiyahetal,GibbonsPope}. As a consequence, the total
Noether charge (\ref{SelfDualNoetherCharge}) vanishes for any isometry, that
is,
\begin{equation}
Q^{\mathrm{NUT}}[\xi ]=0\,,
\end{equation}%
as the electric mass is equal to the magnetic mass. This solution can be
regarded as a family of ground states labeled by $N$.

On the other hand, the Taub-Bolt AdS solution is found for $%
r=r_{b}>\left\vert n\right\vert $ and $f(r=r_{b})=0$. In this case, the
electric mass is
\begin{eqnarray}
Q_{4}^{\mathrm{Bolt}}[\partial _{\tau }] &=&M_{\mathrm{Bolt}}  \notag \\
&=&\frac{r_{b}^{2}+n^{2}}{2r_{b}}-\frac{3}{2\ell ^{2}}\left( \frac{n^{4}}{%
r_{b}}+2n^{2}r_{b}-\frac{r_{b}^{3}}{3}\right) \,.
\end{eqnarray}%
In turn, the magnetic mass for the Bolt solution remains the same as in the
NUT case, such that the total mass and angular momentum are
\begin{eqnarray}
Q_{\mathrm{Bolt}}[\partial _{\tau }] &=&M_{\mathrm{Bolt}}\pm M_{\mathrm{NUT}%
}\,,  \label{MassBolt} \\
Q_{\mathrm{Bolt}}[\partial _{\phi }] &=&0\,.  \label{JBolt}
\end{eqnarray}%
The anti self-dual case in Eq.(\ref{MassBolt}) corresponds to the mass
calculated in Ref.\cite{Hawking:1998ct} following a background-dependent
procedure.

\section{Conclusions}

A fully-covariant expression for the conserved quantities for 4D AdS gravity
supplemented by Gauss-Bonnet and Pontryagin terms has been obtained \`{a} la
Wald.

By analogy with electromagnetism, all the charges are identically zero for
globally self-dual solutions.

A similar expression for the Noether charges has been worked out in Refs.%
\cite{Remi1,Remi2} in first-order formalism. In this Riemann-Cartan
approach, the parity-violating sector appears enlarged by the Holst and
Nieh-Yang terms, which are identically vanishing in Riemannian gravity \cite%
{Corichi et al}. As a consequence, a contribution associated to this new
topological invariant enters in the expression of the Noether charges with
an arbitrary coupling. For a such a case, the surface term is not
proportional to the dual of the Weyl tensor, which implies that no
considerations about the self-duality condition can be made.

We understand that, in Riemann-Cartan theory, a sensible choice of the Holst
coupling is the one that produces the dual of the Weyl tensor at the
boundary for asymptotically AdS spaces, in a similar fashion that only for
the Gauss-Bonnet coupling in Eq.(\ref{I4}) the surface term is proportional
to the Weyl tensor \cite{Carlip}.

Implications of the addition of Pontryagin term and self-duality condition
for the Weyl tensor at the level of the Euclidean action and thermodynamics
of AAdS gravitational objects will be discussed elsewhere.

\section*{Acknowledgements}

The authors thank R. Durka and E. Frodden for interesting discussions. This
work was funded by FONDECYT Grants No. 1131075, 1140296 and 1151107. O.M.
thanks DII-PUCV for support through the project No. 123.736/2015. The work
of R.A. and R.O. is financed in part by the UNAB grants DI-735-15/R and No.
DI-551-14/R, respectively.

\section*{Appendices}

\appendix

\section{Noether theorem \ \ \label{Noether}}

Noether theorem states that for any action invariant under a continuous
transformation, there is a conserved current which leads to a conserved
charge.

Let $I[\phi ]=\int d^{4}x\,L\mathcal{(}\phi ,\partial \phi \mathcal{)}$ be
an action for a set of the fields $\phi (x)$, where the Lagrangian $L$ may
contain boundary terms added to the action. By varying the form of fields, $%
\delta \phi (x)=\phi ^{\prime }(x)-\phi (x)$, an extremum on the action is
reached for the Euler-Lagrange equations,%
\begin{equation}
\frac{\delta I[\phi ]}{\delta \phi }=\frac{\partial L}{\partial \phi }%
-\partial _{\mu }\frac{\partial L}{\partial \partial _{\mu }\phi }=0\,.
\label{EL}
\end{equation}%
The surface term in a general variation of the action%
\begin{equation}
\delta I[\phi ]=\text{e.o.m.}+\int d^{4}x\,\partial _{\mu }\left( \frac{%
\partial L}{\partial \partial _{\mu }\phi }\,\delta \phi \right) \equiv \int
d^{4}x\,\partial _{\mu }\Theta ^{\mu }(\phi ,\delta \phi )\,,  \label{theta}
\end{equation}%
must vanish upon suitable boundary conditions on the field $\phi $, in order
to have a well-posed action principle.

Let us assume that the action $I[\phi ]$ is invariant under the continuous
transformations%
\begin{eqnarray}
x^{\mu } &\rightarrow &x^{\prime \mu }=x^{\mu }+\delta x^{\mu }\,,  \notag \\
\phi (x) &\rightarrow &\phi ^{\prime }(x^{\prime })=\phi (x)+\delta _{T}\phi
(x)\,,  \label{transformations}
\end{eqnarray}%
where the variation of the form of the field, $\delta \phi $, is related to
the total variation of the field, $\delta _{T}\phi $, as%
\begin{equation}
\delta _{T}\phi (x)=\delta \phi (x)+\partial _{\mu }\phi \,\delta x^{\mu }\,.
\label{deltaT_phi}
\end{equation}%
Transformations (\ref{transformations}) are a symmetry of the theory if the
action is off-shell invariant,%
\begin{equation}
\delta I[\phi ]=\int d^{4}x^{\prime }\text{\thinspace }L^{\prime }(x^{\prime
})\,-\int d^{4}x\,L(x)=0\,.  \label{invariance}
\end{equation}

The Noether current is obtained by rewriting the invariance condition (\ref%
{invariance}), and identifying the equations of motion. Using the
Euler-Lagrange equations (\ref{EL}), the Lagrangian changes as
\begin{equation}
\delta L=\frac{\partial L}{\partial \phi }\,\delta \phi +\frac{\partial L}{%
\partial \partial _{\mu }\phi }\,\partial _{\mu }\delta \phi =\partial _{\mu
}\left( \frac{\partial L}{\partial \partial _{\mu }\phi }\,\delta \phi
\right) =\partial _{\mu }\Theta ^{\mu }(\phi ,\delta \phi )\,,
\end{equation}%
and the volume element changes by the Jacobian, $\left\vert \frac{\partial
x^{\prime }}{\partial x}\right\vert \approx 1+\partial _{\mu }\delta x^{\mu
} $. Therefore, the total change in the Lagrangian is%
\begin{equation}
L^{\prime }(x^{\prime })=L(x)+\partial _{\mu }\Theta ^{\mu }(\phi ,\delta
\phi )+\partial _{\mu }L\,\delta x^{\mu }\,.  \label{delta_L}
\end{equation}%
The relations (\ref{deltaT_phi}--\ref{delta_L}) imply that the symmetry
transformations change the action as a total derivative,
\begin{equation}
\delta I[\phi ]=\int d^{4}x\,\partial _{\mu }\left( \Theta ^{\mu }(\phi
,\delta \phi )+L\,\delta x^{\mu }\right) =\int d^{4}x\,\partial _{\mu }(%
\sqrt{g}J^{\mu })\,.
\end{equation}%
Furthermore, the invariance condition (\ref{invariance}) leads to the
conservation law%
\begin{equation}
\partial _{\mu }\left( \sqrt{g}J^{\mu }\right) =0\,.
\end{equation}%
The Noether current is then given by%
\begin{equation}
\sqrt{g}J^{\mu }=\Theta ^{\mu }(\phi ,\delta \phi )+L\,\delta x^{\mu }\,.
\label{J}
\end{equation}%
On the contrary to the situation described in Eq.(\ref{invariance}), if the
action does change by a boundary term $\int d^{4}x\,\partial _{\mu }(\sqrt{g}%
\Omega ^{\mu })$, the conserved current is modified as $\tilde{J}^{\mu
}=J^{\mu }-\Omega ^{\mu }$.

The computation of the conserved charge requires to specify the boundary.
The spacetime has topology $M\simeq \mathbb{R}\times \Sigma $, where $\Sigma
$ is the spatial section with a unit normal vector $u_{\mu }=(-\tilde{N}%
,0,0,0)$. To define invariant volume element, we use the relation $\sqrt{g}=N%
\sqrt{h}=N\tilde{N}\sqrt{\sigma }$.\ The conserved charge reads%
\begin{equation}
Q=\int\limits_{\Sigma }d^{3}x\sqrt{\sigma }N\,u_{\mu }\,J^{\mu }\,.
\end{equation}

If, in turn, the Noether current can be written as a total derivative,
\begin{equation}
\sqrt{g}J^{\mu }=\partial _{\nu }\left( \sqrt{h}Q^{\mu \nu }\right) \,,
\label{J=dQ}
\end{equation}%
then the charge becomes%
\begin{equation}
Q=\int\limits_{\partial \Sigma }d^{2}x\sqrt{\sigma }\,u_{\mu }n_{\nu
}\,Q^{\mu \nu }\,.  \label{Q}
\end{equation}%
Here, $n_{\mu }=(0,N,0,0)$ is a normal to the boundary $\partial M\simeq
\mathbb{R}\times \partial \Sigma $. The quantity $d^{2}x\sqrt{\sigma }%
\,n_{\mu }u_{\nu }$ is the\ dual surface element of $\partial \Sigma $ that
is antisymmetric, such that $Q^{\mu \nu }=-Q^{\nu \mu }$.

\subparagraph{Electromagnetic charge.}

Maxwell electrodynamics with the Pontryagin term is invariant under $U(1)$
gauge transformations, $\delta _{\lambda }A_{\nu }=\partial _{\nu }\lambda $%
. This implies $\delta _{\lambda }F_{\mu \nu }=0$, so that the invariance
condition (\ref{invariance}) of the action is fulfilled. This is an internal
symmetry ($\delta x^{\mu }=0$), and the Noether current (\ref{J}) reads
\begin{equation}
J^{\mu }=\frac{\partial L}{\partial \partial _{\mu }A_{\nu }}\,\partial
_{\nu }\lambda \,.
\end{equation}%
We take $\sigma =1$. Differentiating\ the Lagrangian $L=\frac{1}{4}%
\,(F^{\alpha \beta }F_{\alpha \beta }+\gamma \,^{\ast }F^{\alpha \beta
}F_{\alpha \beta })$ leads to
\begin{eqnarray}
J^{\mu } &=&\left( F^{\mu \nu }+\gamma \,^{\ast }F^{\mu \nu }\right)
\,\partial _{\nu }\lambda  \notag \\
&=&\partial _{\nu }\left[ \left( F^{\mu \nu }+\gamma \,^{\ast }F^{\mu \nu
}\right) \lambda \right] \,,
\end{eqnarray}%
where the last line is obtained using the Maxwell equations and the Bianchi
identity in order to obtain the charge tensor (\ref{J=dQ}) as%
\begin{equation}
Q^{\mu \nu }=\left( F^{\mu \nu }+\gamma \,^{\ast }F^{\mu \nu }\right)
\,\lambda \,.
\end{equation}%
In spherical coordinates, the boundary manifold $\partial M=\mathbb{R}\times
S^{2}$ has a radial normal $n_{\mu }=\delta _{\mu }^{r}$ and the timelike
normal $u_{\nu }=-\delta _{\nu }^{t}$, and the parameter $\lambda $ is
constant on $\partial \Sigma $, such that it can be set to $1$. This enables
to compute the electromagnetic charge as in Eq.(\ref{Q}).

\subparagraph{Diffeomorphic current.}

An action for Riemmanian gravity, with the metric as the only fundamental
field, is invariant under an infinitesimal change of coordinates $\delta x^{\mu
}=\xi ^{\mu }(x)$, where the metric transforms as a Lie derivative,
\begin{equation}
\delta _{\xi }g_{\mu \nu }=-\pounds _{\xi }g_{\mu \nu }=-\left( \nabla _{\mu
}\xi _{\nu }+\nabla _{\nu }\xi _{\mu }\right) \,.
\end{equation}

Since the action depends on $g_{\mu \nu }$ and its derivatives combined in
the Cristoffel symbol $\Gamma _{\alpha \beta }^{\lambda }$, it is convenient
to separate the boundary term (\ref{theta}) which depends on $\delta g_{\mu
\nu }$ from the one that depends on $\delta \Gamma _{\alpha \beta }^{\lambda
} $, so that the Noether current (\ref{J}) can be written as%
\begin{equation}
\sqrt{g}J^{\mu }=\Theta ^{\mu }(g,\delta _{\xi }\Gamma )+\Theta ^{\mu
}(g,\delta _{\xi }g)+L\,\xi ^{\mu }\,.
\end{equation}%
Note that $\Theta ^{\mu }(g,\delta _{\xi }g)=0$ as a consequence of the
asymptotic Killing equation, $\pounds _{\xi }g_{\mu \nu }=\nabla _{\mu }\xi
_{\nu }+\nabla _{\nu }\xi _{\mu }=0$, which describes isometries of the
spacetime.

\section{Asymptotically AdS Spacetimes \ \ \ \label{AAdS}}

We first consider a radial foliation of the spacetime, given by the normal
coordinates
\begin{equation}
ds^{2}=N^{2}(\rho )\,d\rho ^{2}+h_{ij}(\rho ,x)\,dx^{i}dx^{j}\,,
\label{normalradial}
\end{equation}%
where $h_{ij}$ is the induced metric on a boundary $\partial M$ defined at $%
\rho =Const$ and parametrized by the coordinate set $\{x^{i}\}$. In this
frame, the only nonvanishing components of the Christoffel symbol are
\begin{eqnarray}
\Gamma _{ij}^{\rho } &=&\frac{1}{N}\,K_{ij}\,,\qquad \Gamma _{\rho
j}^{i}=-NK_{j}^{i}\,,  \notag \\
\Gamma _{\rho \rho }^{\rho } &=&\frac{d\left( \ln N\right) }{dr}\,,\qquad
\Gamma _{jl}^{i}(g)=\Gamma _{jl}^{i}(h)\,,
\end{eqnarray}%
where $K_{ij}=-\frac{1}{2N}\partial _{\rho }h_{ij}$ is the extrinsic
curvature.

This spacetime foliation implies the Gauss-Codazzi relations
\begin{eqnarray}
R_{jl}^{i\rho } &=&\frac{1}{N}\,(\nabla _{l}K_{j}^{i}-\nabla
_{j}K_{l}^{i})\,,  \notag \\
R_{j\rho }^{i\rho } &=&\frac{1}{N}\,(K_{j}^{i})^{\prime
}-K_{n}^{i}K_{j}^{n}\,,  \notag \\
R_{jl}^{ik} &=&\mathcal{R}%
_{jl}^{ik}(h)-K_{j}^{i}K_{l}^{k}+K_{l}^{i}K_{j}^{k}\,,
\end{eqnarray}%
where $\nabla _{j}=\nabla _{j}(h)$ is the covariant derivative defined with
respect to the boundary metric and $\mathcal{R}_{jl}^{ik}(h)$ is the
intrinsic curvature of $\partial M$.

\subsection{Asymptotic fall-off of boundary tensors}

A suitable choice of the the lapse function and induced metric in Eq.(\ref%
{normalradial}) as $N=\frac{\ell }{2\rho }$ and $h_{ij}=\frac{1}{\rho }%
\,g_{ij}$, that is,
\begin{equation}
ds^{2}=\frac{\ell ^{2}}{4\rho ^{2}}\,d\rho ^{2}+\frac{1}{\rho }%
\,g_{ij\,}dx^{i}dx^{j}\,,
\end{equation}%
makes it easier to work out an asymptotic (Fefferman-Graham (FG)) form of the
boundary fields for AAdS spaces \cite{Fefferman-Graham}. The metric defined
at the asymptotic boundary ($\rho =0$), can be seen as a power-series
expansion. In particular, in four spacetime dimensions,
\begin{equation}
g_{ij}(x,\rho )=g_{(0)ij}(x)+\rho \,g_{(1)ij}(x)+\rho ^{3/2}\,g_{(3/2)ij}(x)+%
\mathcal{O}(\rho ^{2})\,.
\end{equation}%
The coefficient $g_{(3/2)ij}$ cannot be determined from the field equations,
as it corresponds to the response to the boundary source $g_{(0)ij}$, i.e.,
it is proportional to the stress tensor. Because of the fact that there is no Weyl
anomaly at the boundary of 4D AdS gravity, $g_{(3/2)ij}$ is traceless.

In FG coordinate frame, the expansion for the relevant boundary quantities
leads to the expression
\begin{equation}
\sqrt{h}=\frac{\sqrt{g}}{\rho ^{3/2}}=\frac{\sqrt{g_{(0)}}}{\rho ^{3/2}}+%
\mathcal{O}\left( \frac{1}{\rho }\right) \,,  \label{sqrth}
\end{equation}%
and, for the extrinsic curvature
\begin{equation}
K_{j}^{i}(h)=\frac{1}{\ell }\,\delta _{j}^{i}-\rho \ell \,S_{j}^{i}(g)+%
\mathcal{O}(\rho ^{2})\,,  \label{Kexp}
\end{equation}%
where $S_{j}^{i}(g)$ is the Schouten tensor defined as
\begin{equation}
S_{j}^{i}(g)=\mathcal{R}_{j}^{i}(g)-\frac{1}{4}\,\delta _{j}^{i}\,\mathcal{R}%
(g)\,,  \label{Sij}
\end{equation}%
in terms of the boundary Ricci tensor and the Ricci scalar.

For the intrinsic curvature, the asymptotic expansion gives%
\begin{equation}
\mathcal{R}_{jl}^{ik}(h)=\rho \mathcal{R}_{jl}^{ik}(g)=\rho \mathcal{R}%
_{jl}^{ik}(g_{(0)})+\mathcal{O}(\rho ^{2})\,,  \label{Rhexp}
\end{equation}%
relation which is also valid for traces of the boundary Riemann tensor. That
means that Eq.(\ref{Kexp}) can be rewritten in terms of curvatures of $h_{ij}$
in the next-to-leading order%
\begin{equation}
K_{j}^{i}(h)=\frac{1}{\ell }\,\delta _{j}^{i}-\ell S_{j}^{i}(h)+\mathcal{O}%
(\rho ^{2})\,.
\end{equation}%
Equipped with the asymptotic form of the tensorial quantities involved, we
can expand the variation of the total action (\ref{I+P4}) and work out the
holographic version of the electric and magnetic parts of the Weyl tensor.

\subsection{Holographic Stress Tensor in AdS$_{4}$ Gravity}

The projection in the radial foliation (\ref{normalradial}) of the variation
of gravity action in Eq. (\ref{deltaI4}) can be written as
\begin{equation}
\delta I_{4}=\frac{\ell ^{2}}{32\pi G}\int\limits_{\partial M}d^{3}x\sqrt{h}%
\,\delta _{\lbrack jmn]}^{[ikl]}\left[ \left( \delta K_{i}^{j}+\frac{1}{2}%
K_{q}^{j}h^{qs}\delta h_{si}\right) W_{kl}^{mn}+Nh^{mq}\delta \Gamma
_{qi}^{j}(h)W_{kl}^{\rho n}\right] \,.  \label{deltaI4normal}
\end{equation}%
The expansion of the Weyl tensor in FG frame up to quadratic order in $\rho $
is given by
\begin{eqnarray}
W_{jl}^{i\rho } &=&\mathcal{O}(\rho ^{2})\,,  \notag \\
W_{j\rho }^{i\rho } &=&-\frac{3}{2}\frac{\rho ^{3/2}}{\ell ^{2}}%
g_{(0)}^{ik}g_{(3/2)kj}+\mathcal{O}(\rho ^{2})\,,  \notag \\
W_{jl}^{ik} &=&\rho \,\mathcal{W}_{jl}^{ik}[g_{(0)}]+\frac{3}{2}\frac{\rho
^{3/2}}{\ell ^{2}}\,g_{(0)}^{[im}g_{(3/2)m[j}\delta _{l]}^{k]}+\mathcal{O}%
(\rho ^{2})\,,
\end{eqnarray}%
where the first term in the last relation is the Weyl tensor of the metric
at the conformal boundary $g_{(0)ij}$.

At the same time, we can see the contribution coming from the quantities
that involve variations in the Eq.(\ref{deltaI4normal}), that is,%
\begin{eqnarray}
\delta K_{i}^{j} &=&\mathcal{O}(\rho )\,,  \notag \\
h^{qs}\delta h_{si} &=&g_{(0)}^{qs}\delta g_{(0)si}+\mathcal{O}(\rho )\,,
\notag \\
\delta \Gamma _{qi}^{j}(h) &=&\mathcal{O}(1)\,.
\end{eqnarray}%
A simple power-counting argument applied to the expansion of Eq.(\ref%
{deltaI4normal}) shows that the first and the last terms are subleading, and
actually go to zero as one approaches the boundary $\rho \rightarrow 0$. As
expected, the finite of the variation of EH action plus GB term is the
holographic stress tensor \cite{Miskovic:2009bm,Balasubramanian-Kraus}
\begin{equation}
\delta I_{4}=\frac{1}{2}\int\limits_{\partial M}d^{3}x\sqrt{g_{(0)}}\left( -%
\frac{3}{16\pi G\ell }\,g_{(0)}^{im}\,g_{(3/2)mn}\,g_{(0)}^{nj}\right)
\delta g_{(0)ij}\,,  \label{deltaI4holo}
\end{equation}%
where we have used the fact that any trace of the boundary Weyl tensor is
zero.

We can covariantize back the above expression in terms of tensorial
quantities related to the full boundary metric $h_{ij}$ and prove that, up
to the relevant order, the variation of $I_{4}$ can be cast in the form
\begin{equation}
\delta I_{4}=-\frac{\ell }{16\pi G}\int\limits_{\partial M}d^{3}x\sqrt{h}%
\,W_{jk}^{ik}\,\left( h^{-1}\delta h\right) _{i}^{j}\,.
\label{deltaI4Wtrace}
\end{equation}%
Using the fact that a single trace of the Weyl tensor is zero, we have that%
\begin{equation}
W_{jk}^{ik}=-W_{j\rho }^{i\rho }\,,
\end{equation}%
and it is easy to show that the quantity that appears at the boundary is the
electric part of the Weyl tensor%
\begin{equation}
E_{j}^{i}=W_{j\nu }^{i\mu }\,n_{\mu }n^{\nu }\,,
\end{equation}%
as we can rewrite Eq.(\ref{deltaI4Wtrace}) in the form%
\begin{equation}
\delta I_{4}=\frac{\ell }{16\pi G}\int\limits_{\partial M}d^{3}x\sqrt{h}%
\,E_{j}^{i}\left( h^{-1}\delta h\right) _{i}^{j}\,.
\end{equation}

This makes manifest the link between the concept of Conformal Mass \cite%
{Ashtekar-Magnon-Das} and the addition of the Gauss-Bonnet term in 4D AdS
gravity \cite{JKMO}.

\subsection{Holographic Cotton Tensor}

The Pontryagin term, written as a boundary term in the coordinate frame (\ref%
{normalradial}), is expressed as
\begin{eqnarray}
\int\limits_{M}d^{4}x\,\mathcal{P}_{4} &=&\int\limits_{\partial M}d^{3}x\,%
\frac{n_{\mu }}{N}\,\epsilon ^{\mu \nu \alpha \beta }\left( \Gamma _{\nu
\lambda }^{\sigma }\partial _{\alpha }\Gamma _{\beta \sigma }^{\lambda }+%
\frac{2}{3}\,\Gamma _{\nu \lambda }^{\sigma }\Gamma _{\alpha \epsilon
}^{\lambda }\Gamma _{\beta \sigma }^{\epsilon }\right)  \notag \\
&=&\int\limits_{\partial M}d^{3}x\,\epsilon ^{ijk}\left[ -\Gamma
_{im}^{l}\left( \partial _{j}\Gamma _{kl}^{m}+\frac{2}{3}\,\Gamma
_{jn}^{m}\Gamma _{kl}^{n}\right) +2K_{i}^{l}\nabla _{j}K_{kl}\right] \,.
\end{eqnarray}%
Using the asymptotic form of the fields in FG expansion, the last term reads%
\begin{equation}
2\epsilon ^{ijk}K_{i}^{l}\nabla _{j}K_{kl}=2\epsilon ^{ijk}\left( \frac{1}{%
\ell }\,\delta _{i}^{l}-\rho \ell S_{i}^{l}+\mathcal{O}(\rho ^{2})\right)
\left( -\ell \nabla _{j}S_{kl}+\mathcal{O}(\rho )\right) \,,
\end{equation}%
in terms of a Schouten tensor and the covariant derivative defined with
respect the conformal metric $g_{(0)ij}\,$. Manipulating the last relation,
we see that%
\begin{equation}
2\epsilon ^{ijk}K_{i}^{l}\nabla _{j}K_{kl}=-2\epsilon ^{ijk}\nabla
_{i}S_{jk}+\mathcal{O}(\rho )=\mathcal{O}(\rho )\,,
\end{equation}%
because $S_{jk}$ is symmetric.

Therefore, using $\delta \Gamma _{im}^{l}=\frac{1}{2}\,h^{ln}\left( \nabla
_{i}\delta h_{nm}+\nabla _{m}\delta h_{ni}-\nabla _{n}\delta h_{im}\right) $%
, the variation of the Pontryagin invariant takes the form,
\begin{eqnarray}
\delta P_{4} &=&-\int\limits_{\partial M}d^{3}x\,\epsilon ^{ijk}\delta
\Gamma _{im}^{l}\,\mathcal{R}_{\ ljk}^{m}(h)  \notag \\
&=&\int\limits_{\partial M}d^{3}x\,\epsilon ^{ijk}\,\left( h^{-1}\delta
h\right) _{i}^{l}\,\nabla _{m}\mathcal{R}_{\ ljk}^{m}\,.
\end{eqnarray}

As the boundary is three-dimensional, its Weyl tensor vanishes,%
\begin{equation}
0=\mathcal{W}_{jk}^{ml}(h)=\mathcal{R}_{jk}^{ml}(h)-\delta
_{j}^{m}S_{k}^{l}(h)+\delta _{j}^{l}S_{k}^{m}(h)+\delta
_{k}^{m}S_{j}^{l}(h)-\delta _{k}^{l}S_{j}^{m}(h)\,,
\end{equation}%
such that%
\begin{equation}
\delta \int\limits_{M}d^{4}x\,\mathcal{P}_{4}=\int\limits_{\partial
M}d^{3}x\,\epsilon ^{ijk}\,\left( h^{-1}\delta h\right) _{i}^{l}\,\nabla
_{m}\left( 2\delta _{j}^{m}S_{kl}-2h_{lj}S_{k}^{m}\right) \,,
\end{equation}%
where the second term in the first line identically vanishes due to the
symmetry of the indices. In doing so, the variation is written as%
\begin{equation}
\delta \int\limits_{M}d^{4}x\,\mathcal{P}_{4}=2\int\limits_{\partial
M}d^{3}x\,\sqrt{h}\,\left( h^{-1}\delta h\right) _{i}^{l}\,C_{l}^{i}\,,
\label{deltaP4final}
\end{equation}%
where $C_{l}^{i}$ is the Cotton-York tensor,%
\begin{equation}
C_{l}^{i}=\frac{1}{\sqrt{h}}\,\epsilon ^{ijk}\,\nabla _{j}S_{kl}\,.
\end{equation}%
Finally, putting together the holographic stress tensor in Eq.(\ref%
{deltaI4holo}) and by rescaling the Cotton tensor in Eq.(\ref%
{deltaP4final}), we see that the finite part of the variation of the total
action is%
\begin{equation}
\delta I=\frac{1}{2}\int\limits_{\partial M}d^{3}x\sqrt{g_{(0)}}\left(
T^{ij}\mp \frac{\ell ^{2}}{8\pi G}\,C^{ij}(g_{(0)})\right) \delta
g_{(0)ij}\,,
\end{equation}%
where $T^{ij}$ is the holographic stress tensor.

In Ref.\cite{Ayan et al}, the holographic reconstruction of gravity is
performed for perfect-Cotton geometries, where the Cotton tensor of the
boundary geometry is proportional to the energy-momentum tensor. A
corresponding gravity theory in the bulk is characterized by self-duality
condition for the Weyl tensor.


\end{document}